\documentclass[prb,preprint,superscriptaddress,showpacs]{revtex4}
\usepackage{amsfonts}
\usepackage{amssymb}
\usepackage{amsmath}
\usepackage{hyperref}
\usepackage{graphicx}

\newcommand{\bfy}{\mathbf{y}}
\newcommand{\bfr}{\mathbf{r}}
\newcommand{\calM}{\mathcal{M}}
\newcommand{\bfysub}{\mathbf{\scriptstyle y}}
\newcommand{\bfrsub}{\mathbf{\scriptstyle r}}
\newcommand{\bfsig}{\mbox{\boldmath $\sigma$}}
\newcommand{\bfehat}{\hat{\mathbf{e}}}

\begin{document}
\title{Density functional theory and quantum computation}

\date{\today}

\author{Frank Gaitan}
\affiliation{Advanced Science Institute, The Institute of Physical and
Chemical Research (RIKEN), Wako-shi, Saitama 351-0198, Japan}
\affiliation{Department of Physics, Southern Illinois University, Carbondale,
IL 62901-4401}
\affiliation{CREST, Japan Science and Technology Agency (JST), Kawaguchi,
Saitama, 332-0012, Japan}

\author{Franco Nori}
\affiliation{Advanced Science Institute, The Institute of Physical and
Chemical Research (RIKEN), Wako-shi, Saitama 351-0198, Japan}
\affiliation{CREST, Japan Science and Technology Agency (JST), Kawaguchi,
Saitama, 332-0012, Japan}
\affiliation{Physics Department, Center for Theoretical Physics, University
of Michigan, Ann Arbor, MI 48109}

\begin{abstract}
This paper establishes the applicability of density functional theory methods 
to quantum computing systems. We show that ground-state and time-dependent 
density functional theory can be applied to quantum computing systems by 
proving the Hohenberg-Kohn and Runge-Gross theorems for a fermionic 
representation of an $N$ qubit system. As a first demonstration of this 
approach, time-dependent density functional theory is used to determine the 
minimum energy gap $\Delta (N)$ arising when the quantum adiabatic evolution 
algorithm is used to solve instances of the NP-Complete problem MAXCUT. It is 
known that the computational efficiency of this algorithm is largely 
determined by the large-$N$ scaling behavior of $\Delta (N)$, and so 
determining this behavior is of fundamental significance. As density 
functional theory has been used to study quantum systems with $N\sim 10^{3}$ 
interacting degrees of freedom, the approach introduced in this paper raises 
the realistic prospect of evaluating the gap $\Delta (N)$ for systems with 
$N\sim 10^{3}$ qubits. Although the calculation of $\Delta (N)$ serves to
illustrate how density functional theory methods can be applied to problems in
quantum computing, the approach has a much broader range, and shows promise as 
a means for determining the properties of very large quantum computing systems.
\end{abstract}

\pacs{31.15.ee,03.67.Ac}

\maketitle

\section{Introduction}
\label{sec1}

The inability of a classical computer to efficiently simulate the dynamics
of a quantum system is well-known. The problem is that the dimension of the 
Hilbert space grows exponentially with the number of degrees of freedom of the
quantum system, and this in turn causes an exponential growth in the amount of 
memory and CPU-time required to carry out the simulation. This inefficiency is 
a major stumbling block for numerical studies aiming to determine the 
asymptotic performance of quantum algorithms. For example, numerical 
simulation of the dynamics of the quantum adiabatic evolution (QAE) algorithm 
applied to the NP-Complete problem \textit{Exact Cover~3\/} has been limited 
to systems containing $N\leq 20$ qubits \cite{farhi,gait}. Because the 
algorithm dynamics must be adiabatic, its runtime $T$ must satisfy the 
inequality
\begin{equation}
T \gg \frac{M}{\Delta^{2}} ;
\end{equation} 
where
\begin{eqnarray}
M & = & \max_{0\leq s\leq 1}\,\left| \langle E_{1}(s)|\frac{dH(s)}{ds}|
          E_{0}(s)\rangle\right| ; \nonumber\\
 & & \nonumber\\
\Delta & = & \min_{0\leq s\leq 1}\,\left[ E_{1}(s)-E_{0}(s)\right] ;
\end{eqnarray}
and here $t$ is time; $s=t/T$ is dimensionless time; $H(s)$ is the 
time-dependent Hamiltonian that drives the dynamics of the QAE algorithm; and 
$\{ E_{i}(s), |E_{i}(s)\rangle : i=0,\ldots ,2^{N}-1\}$ are the eigenvalues 
and eigenstates of $H(s)$. In the usual formulation\cite{farhi2,nori1,nori2} 
of QAE, 
$dH(s)/ds$ is an $s$-independent matrix whose largest eigenvalue bounds $M$. 
Typically, this eigenvalue scales polynomially with $N$. Thus, if the minimum 
gap $\Delta (N)$ separating the ground- and first-excited states scales 
polynomially (exponentially) with $N$, so will the algorithm runtime $T(N)$. 
An efficient (inefficient) algorithm\cite{gar&john} for a computational
problem is one that solves all instances of the problem with polynomial 
(exponential) $T(N)$. We see then that the computational efficiency of the 
QAE algorithm is largely determined by the scaling behavior of the minimum gap 
$\Delta (N)$. Attempts to evaluate $\Delta (N)$ using exact 
diagonalization\cite{hogg} have been limited to $N \leq 20$ qubits. Recently, 
however, the minimum gap $\Delta (N)$ for QAE applied to Exact Cover~3 has 
been determined for $N \leq 128$ qubits using quantum Monte Carlo 
methods\cite{apy}. This represents a substantial technical advance, and has 
stirred great interest in finding other computational approaches that might 
allow quantum algorithm performance to be determined for still larger qubit 
systems.

Quantum computation is not the only research area struggling with the 
difficulties of simulating quantum systems\cite{bul&nori}. Condensed-matter 
physicists and 
quantum chemists have been working under the shadow of this problem for 
decades. A number of computational approaches have been developed which, 
together with increasingly more powerful computers, have allowed much progress 
to be made, in spite of the ultimately unavoidable difficulties involved. Among 
these approaches, density functional theory (DFT) has proven to be one of the 
most successful \cite{d&g,p&w,marq}. DFT is a theory of \textit{interacting\/} 
fermion systems. It provides an \textit{exact\/} treatment of all many-body 
effects through the exchange-correlation energy functional. It can also handle
the coupling of such fermion systems to both static and time-varying electric 
and magnetic fields. Ground-state density functional theory (GS-DFT) has been 
used to determine a wide range of ground-state properties of atomic, 
molecular, and solid state systems \cite{hk,ks}; while time-dependent 
density functional theory (TD-DFT) has been used to determine excited-state 
properties, as well as the linear and non-linear response of interacting 
many-electron systems to electromagnetic fields \cite{r&g,m&g}. For our 
purposes, it is especially significant that DFT has been successfully applied 
to quantum systems containing $N\sim 10^{3}$ interacting degrees of freedom 
\cite{shimo,jiang,sanch}. 

In this paper we establish the applicability of DFT methods to quantum 
computing systems. By establishing this link, we shall see that
a powerful tool becomes available for determining the properties of very large 
quantum computing systems. Although our analysis can be extended to the case 
of $N$ qudits ($d$-level systems) residing on a $D$-dimensional lattice, we 
restrict the presentation to $N$ qubits residing on a 2D lattice since this 
corresponds to the experimentally interesting cases of qubits placed in a 2D 
ion trap\cite{chen}, or restricted to a planar superconducting qubit 
circuit\cite{you}. 

The outline of this paper is as follows. We begin in Section~\ref{sec2} by 
showing how an $N$ qubit system can be transformed into a system of $N$ 
lattice fermions, and then in Section~\ref{sec3}, illustrate this 
transformation by using it to re-write the dynamics of the QAE algorithm 
applied to the NP-Complete problem \textit{MAXCUT}\cite{stef}. For the 
resulting interacting fermion system, Section~\ref{sec4} establishes the 
Hohenberg-Kohn \cite{hk} and Runge-Gross \cite{r&g} theorems, and sets up the 
auxiliary Kohn-Sham system of non-interacting fermions \cite{ks}. The results 
of Section~\ref{sec4} provide the justification for applying GS- and TD-DFT to 
quantum computing systems. The proofs given in Section~\ref{sec4} are 
adaptations of well-established proofs used for interacting electron systems,
and so their validity should not be in doubt. Section~\ref{sec5} works out the 
linear response of the system of interacting fermionized qubits using TD-DFT, 
and as an application, shows how this response can be used to determine the 
minimum energy gap $\Delta (N)$ for the MAXCUT dynamics. Here we begin to see 
the value of the newly established link between DFT and quantum computing. 
Calculation of $\Delta (N)$ boils down to a calculation of excitation 
energies, and the reliable calculation of excitation energies for very large 
interacting electron systems was one of the first triumphs of TD-DFT. A
straightforward adaptation of standard TD-DFT arguments then determines 
$\Delta (N)$. In light of earlier remarks, the link established in this paper
between DFT and quantum computing raises the realistic prospect of evaluating 
the minimum gap $\Delta (N)$ for $N \sim 10^{3}$ qubits, and thus of studying
the performance of the QAE algorithm for much larger qubit systems than is 
currently possible using other approaches. Although we focus on the calculation
of the minimum gap in this paper, it is clear that the application of DFT to 
quantum computing systems has a much broader range, and shows genuine promise 
as a means for determining the properties of very large quantum computing 
systems. Finally, the paper closes in Section~\ref{sec6} with a discussion of 
future work.

\section{Qubit--Fermion Transformation} 
\label{sec2}

Consider $N$ qubits 
residing on an $N$-site 2D lattice with basis vectors $\hat{\mathbf{e}}_{k}$ 
$(k=1,2)$, and sites specified by the position vector $\bfr$. Let 
$\bfsig (\bfr)$ denote the Pauli matrices associated with the qubit at $\bfr$. 
We now show how the qubits can be converted into lattice fermions via the 2D 
Jordan-Wigner (JW) transformation \cite{frad}. Note that the following 
analysis can be extended to $N$ qudits (viz.~$N$ $d$-level systems) on a 
$D$-dimensional lattice using the generalized JW transformation \cite{b&o} 
that fermionizes a spin $s$ system ($d=2s+1$) in $D$ spatial dimensions. 
 
For a 2D system of qubits, the JW transformation is:
\begin{eqnarray}
   \sigma^{+}(\bfr)  & = & 2a^{\dag}_{\bfrsub} Q_{\bfrsub}  \nonumber\\ 
   \sigma^{-}(\bfr) & = &  2Q^{\dag}_{\bfrsub} a_{\bfrsub}   \nonumber\\  
   \sigma_{z}(\bfr)  & = &  2\hat{n}_{\bfrsub}  -1 . 
\label{allequationsJW}
\end{eqnarray}
Here: $\sigma^{\pm}(\bfr ) = \sigma_{x}(\bfr ) \pm i\sigma_{y}(\bfr )$;
$a^{\dag}_{\bfrsub}$ ($a_{\bfrsub}$) creates (annihilates) a lattice fermion 
at $\bfr$; $\hat{n}_{\bfrsub} = a^{\dag}_{\bfrsub}a_{\bfrsub}$ is the
fermion number operator at $\bfr$; and
\begin{eqnarray}
Q_{\bfrsub} = \exp \left[ -i \phi_{\bfrsub}\right] 
   & \hspace{0.025in} ; \hspace{0.025in} &
\phi_{\bfrsub} = \left(\frac{1}{2\pi\theta}\right)
\sum_{\bfr^{\prime}}\,\Phi (\bfr , \bfr^{\prime})\,\hat{n}_{\bfrsub^{\prime}}. 
\label{disorder}
\end{eqnarray}
In Eq.~(\ref{disorder}), $\Phi (\bfr ,\bfr^{\prime})$ is the angle made by 
$(\bfr -\bfr^{\prime})$ with respect to some reference direction, say 
$\hat{\mathbf{e}}_{1}$. Thus: (i)~$\Phi (\bfr ,\bfr^{\prime})$ changes by 
$2\pi$ when $(\bfr -\bfr^{\prime})$ traces out a closed loop around 
$\bfr^{\prime}$; and (ii)~by convention, $\Phi (\bfr , \bfr ) \equiv 0$. 
The requirement that the Pauli operators $\bfsig (\bfr )$ commute at different 
lattice sites forces $\theta$ to satisfy 
\begin{displaymath}
\frac{1}{2\pi\theta}  = 2m + 1 \hspace{0.25in} (m=0,\pm 1,\pm 2, \cdots ) 
\end{displaymath} 
in Eq.~(\ref{disorder}).

As shown in Ref.~\onlinecite{frad}, the lattice fermions are spinless, and 
minimally-coupled to a gauge field $A_{k}(\bfr ) = \Delta_{k}\phi_{\bfrsub} 
\equiv \phi_{\bfrsub + \bfehat_{k}} -\phi_{\bfrsub}$. The action for the 
gauge-field $A_{\mu}(\bfr )$ is given by the Chern-Simons term\cite{foot}
\begin{displaymath}
\mathcal{A}_{cs} = -\frac{\theta}{4}\int dt\sum_{\bfrsub}
                     \epsilon^{\mu\nu\lambda}A_{\mu}(\bfr )
                        F_{\nu\lambda}(\bfr ) .
\end{displaymath}
Maxwell's equations for this system take the form 
\begin{equation}
j^{\mu}_{\bfrsub} = 
\epsilon^{\mu\nu\lambda}F_{\nu\lambda}(\bfr ), 
\label{jFrel}
\end{equation}
where $j^{\mu}_{\bfrsub}$ is 
the fermion current, $F_{\nu\lambda}(\bfr )$ is the gauge field tensor, 
$\epsilon^{\mu\nu\lambda}$ is the totally anti-symmetric Levi-Civita tensor, 
and $\mu ,\nu , \lambda = 0,1,2$. From Eq.~(\ref{jFrel}), the fermion current 
$j_{\bfrsub ,\mu}$ has components
\begin{eqnarray}
j_{\bfrsub ,0} & = & n_{\bfrsub} , \nonumber\\
j_{\bfrsub ,k} & = & \frac{1}{2\pi}\sum_{\bfysub}\left\{ \Delta_{k}
                      G_{\bfrsub ,\bfysub}\right\} 
                       \left\{\partial_{t}n_{\bfysub}\right\}
        \hspace{0.2in} (k=1,2),
\label{fermcurr}
\end{eqnarray}
where $G_{\bfrsub ,\bfysub}$ is the Green's function for the lattice Laplacian
\begin{displaymath}
\sum_{k=1,2}\Delta_{k}\Delta_{k} G_{\bfrsub ,\bfysub} = -2\pi
    \delta_{\bfrsub ,\bfysub} .
\end{displaymath}
Fermion current conservation, $\partial_{\mu}j^{\mu}=0$, follows immediately 
from Maxwell's equations.

\section{Application: NP-Complete Problem MAXCUT} 
\label{sec3}

In the problem MAXCUT, one considers an $N$-node undirected graph $\mathcal{G}$
with nodes specified by $\bfr$. The nodes (edges) are assigned weights
$w_{\bfrsub}$ ($w_{\bfrsub ,\bfrsub^{\prime}}$), and a binary variable 
$s_{\bfrsub}$ is associated with each node $\bfr$. A cut of the graph 
$\mathcal{G}$ is a partition of the nodes into two sets $\mathcal{S}_{0}$ and 
$\mathcal{S}_{1}$. For all nodes belonging to $\mathcal{S}_{0}$ 
($\mathcal{S}_{1}$), $s_{\bfrsub}$ is assigned the value $0$ ($1$). The node 
variables are used to construct a string variable $s=s_{\bfrsub_{1}}\cdots 
s_{\bfrsub_{N}}$, and all possible assignments of the $N$ (binary) node 
variables leads to $2^{N}$ possible string assignments for $s$. It follows that
there is a one-to-one correspondence between cuts of $\mathcal{G}$ and string 
assignments for $s$. The MAXCUT problem is to find the cut (viz.~string 
assignment) that maximizes the payoff function $P(s)$ given by
\begin{displaymath}
P(s) =\sum_{\bfr} s_{\bfrsub} w_{\bfrsub} + \sum_{\bfr ,
\bfr^{\prime}} s_{\bfrsub}(1-s_{\bfrsub^{\prime}}) w_{\bfrsub ,
\bfrsub^{\prime}} .
\end{displaymath} 
MAXCUT is known\cite{gar&john} to be NP-Complete, and so it belongs to the set 
of ``hardest problems'' in the complexity class NP.  

The QAE algorithm was applied to MAXCUT in Ref.~\onlinecite{stef}, where the 
dynamics is driven by the Hamiltonian 
\begin{equation}
H(t) = \left( 1-\frac{t}{T}\right) H_{0} + \left(\frac{t}{T}\right) H_{P}.
\label{QAEHam}
\end{equation}
Here $T$ is the algorithm runtime, 
\begin{equation}
H_{0}  =  \sum_{\bfr}\sigma_{x}(\bfr ), 
\label{H0def}
\end{equation}
and
\begin{eqnarray}
H_{P}  & = & \sum_{\bfr}w_{\bfrsub}( 1-\sigma_{z}(\bfr ))/2 \nonumber\\
  & & {}\hspace{0.60in} + \sum_{\bfr , \bfr^{\prime}} 
         w_{\bfrsub , \bfrsub^{\prime}}( 1 - 
       \sigma_{z}(\bfr ) \sigma_{z}(\bfr^{\prime}))/2 . 
\label{HPMAX}
\end{eqnarray}
The Hamiltonian $H_{P}$ is known as the problem Hamiltonian. From
Eq.~(\ref{HPMAX}), its eigenstates are the simultaneous eigenstates of the 
$\{\sigma_{z}(\bfr )\}$: 
\begin{displaymath}
\sigma_{z}(\bfr )|s_{\bfrsub_{1}}\cdots 
s_{\bfrsub_{N}}\rangle = (-1)^{s_{\protect\scriptscriptstyle\bfrsub}}
  |s_{\bfrsub_{1}}\cdots 
s_{\bfrsub_{N}}\rangle. 
\end{displaymath}
\textit{By construction\/}\cite{stef,farhi}, each bit string 
$s=s_{1}\cdots s_{N}$ that maximizes the MAXCUT payoff function labels a 
ground-state $|s_{1}\cdots s_{N}\rangle$ of $H_{P}$. The QAE algorithm places 
the qubit system in the ground-state of the initial Hamiltonian $H_{0}$, and 
for runtime $T$ sufficiently large, $H(t)$ evolves the quantum state 
adiabatically so that at time $T$, the system is in the ground-state of the 
final Hamiltonian $H_{P}$ with probability approaching $1$. Measurement of the 
$\{\sigma_{z}(\bfr )\}$ at time $T$ yields, with probability approaching $1$, 
a string $s_{1}\cdots s_{N}$ that solves the MAXCUT instance.

Using Eqs.~(\ref{allequationsJW}) in $H(t)$ gives the fermionized QAE 
Hamiltonian for MAXCUT:
\begin{eqnarray}
{}\hspace{-0.25in}H(t) & = & \left( 1-\frac{t}{T}\right)
                   \sum_{\bfr}[ a^{\dag}_{\bfrsub}
                              Q_{\bfrsub}+ Q^{\dag}_{\bfrsub} a_{\bfrsub}]
               \nonumber\\
 & & {} \hspace{-0.15in}+\left(\frac{t}{T}\right)
          \sum_{\bfr} v_{\bfrsub}\hat{n}_{\bfrsub} 
      + \left(\frac{t}{T}\right)\sum_{\bfr}\sum_{\bfr^{\prime}\neq\bfr} 
         w_{\bfrsub ,\bfrsub^{\prime}}\hat{n}_{\bfrsub}
          \hat{n}_{\bfrsub^{\prime}}] ,
\label{fermHam}
\end{eqnarray} 
where 
\begin{displaymath}
w_{\bfrsub}  \equiv  v_{\bfrsub} + W_{\bfrsub}; 
\end{displaymath}
\begin{displaymath}
W_{\bfrsub} = 
\sum_{\bfr^{\prime}\neq \bfr}w_{\bfrsub ,\bfrsub^{\prime}};
\end{displaymath} 
and a term proportional to the identity has been suppressed. 

\section{Density Functional Theory} 
\label{sec4}

In this Section we establish the applicability of the Hohenberg-Kohn and 
Runge-Gross theorems to the QAE/MAXCUT problem. These theorems justify the 
use of, respectively, ground-state and time-dependent density functional 
theory to the MAXCUT dynamics. Throughout, the ground-state is assumed to
be non-degenerate, as would be appropriate for a non-vanishing minimum gap
$\Delta$. The formalism can be extended, however, to cover degenerate 
ground-states\cite{kohn,dr&gro}.

\subsection{Ground-State Density Functional Theory}
\label{sec4.1}

We have seen that the QAE algorithm has an adiabatic dynamics that is driven
by a slowly-varying Hamiltonian $H(t)$. In this subsection we focus on the 
MAXCUT Hamiltonian $H(t)$ at a \textit{fixed instant of time\/} $t=t_{\ast}$. 
By fixing the time, we obtain a well-defined \textit{static\/} Hermitian 
operator $H_{\ast}\equiv H(t=t_{\ast})$. The aim of this subsection is to show 
that the Hohenberg-Kohn theorem applies to $H_{\ast}$. With this theorem in 
place, GS-DFT can be used to study the ground-state properties of $H_{\ast}=
H(t=t_{\ast})$ for any specific intermediate time $0<t_{\ast}<T$. We stress 
that even though the QAE algorithm works with a slowly-varying Hamiltonian, 
the discussion in this subsection is restricted to the \textit{static\/} 
Hermitian operator $H_{\ast}= H(t_{\ast})$ that is the value of $H(t)$ at the 
time $t=t_{\ast}$. 

Our starting point is the energy functional for the instantaneous MAXCUT 
Hamiltonian $H_{\ast}\equiv H(t=t_{\ast})$:
\begin{equation}
E[n] = \min_{|\psi\rangle\rightarrow n}\langle\psi |H_{\ast}|\psi\rangle .
\label{enfunc}
\end{equation} 
The domain of $E[n]$ is the set of all $N$-representable site occupation
functions (SOF) $n_{\bfrsub}$ that can be obtained from an $N$-fermion 
wavefunction. The minimization in Eq.~(\ref{enfunc}) is over all $|\psi
\rangle$ for which 
\begin{displaymath}
n\equiv n_{\bfrsub}=\langle\psi |\hat{n}_{\bfrsub}| \psi
\rangle, 
\end{displaymath}
and the minimizing state $|\psi_{\min}[n] \rangle$ is thus a functional 
of $n_{\bfrsub}$. Let $|\psi^{g}\rangle$ denote the ground-state of 
$H_{\ast}$; $E^{g}$ the ground-state energy; and $n^{g}_{\bfrsub}$ the 
ground-state SOF. Inserting Eq.~(\ref{fermHam}) into (\ref{enfunc}) gives 
\begin{displaymath}
E[n] = (t_{\ast}/T)\sum_{\bfrsub}v_{\bfrsub}n_{\bfrsub} + Q[n],
\end{displaymath}
where 
\begin{displaymath}
Q[n]\equiv\min_{|\psi\rangle\rightarrow n}\langle\psi |(T_{t_{\ast}}+
   U_{t_{\ast}})|
\psi\rangle,
\end{displaymath} 
and $T_{t_{\ast}}$ and $U_{t_{\ast}}$ are the first and third terms,
respectively, on the RHS of Eq.~(\ref{fermHam}) at $t=t_{\ast}$. 

To establish the Hohenberg-Kohn (HK) theorem for $H_{\ast}$ we must show 
\cite{levy}: (i)~$E[n^{g}]=E^{g}$; (ii)~for 
$n_{\bfrsub}\neq n^{g}_{\bfrsub}$, $E[n]>E^{g}$; and (iii)~the ground-state 
expectation value of any observable is a unique functional of the ground-state 
SOF $n^{g}_{\bfrsub}$. By the variational principle, $\langle\psi |H_{\ast}
|\psi\rangle\geq E^{g}$, with equality when $|\psi\rangle = |\psi^{g}\rangle$. 
Thus, for $n=n^{g}$, the search in Eq.~(\ref{enfunc}) returns the ground-state
$|\psi^{g}\rangle$ as the state $|\psi_{\min}[n^{g}]\rangle$ that minimizes 
$E[n^{g}]$. It follows that 
\begin{displaymath}
E[n^{g}]=\langle\psi^{g}|H_{\ast}| \psi^{g}\rangle = 
E^{g}. 
\end{displaymath}
This establishes condition~(i). For $n\neq n^{g}$, the minimizing 
state $|\psi_{\min}[n]\rangle\neq |\psi^{g} [n^{g}]\rangle$, and so by the 
variational principle, 
\begin{displaymath}
E[n]=\langle\psi_{\min} [n]|H_{\ast}|\psi_{\min}[n]
\, \rangle\; > \; E^{g}. 
\end{displaymath}
This establishes condition~(ii). Finally, since the 
ground-state $|\psi^{g}\rangle = |\psi_{\min}[n^{g}] \rangle$, it is a 
functional of $n^{g}$, and consequently, so are all ground-state expectation 
values: 
\begin{eqnarray*}
\langle\hat{\mathcal{O}}\rangle_{gs} & = & \langle\psi^{g}|\hat{\mathcal{O}}
                                            |\psi^{g}\rangle \\
  & = & \langle \psi_{\min}[n^{g}] |\:\hat{\mathcal{O}}\: |\psi_{\min}[n^{g}] 
              \,\rangle \\
  & = & \mathcal{O}[n^{g}].
\end{eqnarray*} 
Condition~(iii) is thus established, completing the proof of the HK theorem 
for $H_{\ast}=H(t_{\ast})$.

To obtain a practical calculational scheme, an auxiliary system of 
non-interacting Kohn-Sham (KS) fermions is introduced\cite{ks}, and it is 
assumed that the ground-state SOF $n^{g}_{\bfrsub}$ can be obtained 
from the ground-state density of the KS fermions moving in an external 
potential $v^{ks}_{\bfrsub}$. For $H_{\ast}=H(t_{\ast})$, the KS Hamiltonian 
$H_{ks}=T^{\prime}_{t_{\ast}}+V^{ks}$ is defined to be:
\begin{displaymath}
H_{ks} = \sum_{\bfrsub}\left( 1-\frac{t_{\ast}}{T}\right) \{ q_{\bfrsub}
             a^{\dagger}_{\bfrsub}
                  + q^{\ast}_{\bfrsub}a_{\bfrsub}\} + 
                \sum_{\bfrsub}\left(\frac{t_{\ast}}{T}\right)
                                    v^{ks}_{\bfrsub}\hat{n}_{\bfrsub} , 
\end{displaymath} 
where $q_{\bfrsub}=\langle Q_{\bfrsub}\rangle$ is the ground-state expectation
value of $Q_{\bfrsub}$. The effects of $Q_{\bfrsub}$ are thus incorporated 
into the KS dynamics through the mean-field $q_{\bfrsub}$. The KS energy 
functional $\epsilon_{ks}[n]$ is:
\begin{equation}
\epsilon_{ks}[n] = \min_{|\psi\rangle\rightarrow n}\langle\psi |H_{ks}|
                       \psi\rangle = T^{\prime}_{t_{\ast}}[n]+\sum_{\bfrsub}
                           \left(\frac{t_{\ast}}{T}\right)
                              v^{ks}_{\bfrsub}n_{\bfrsub} .
\label{ksenfunc}
\end{equation}
To determine the KS external potential $v^{ks}_{\bfrsub}$, we re-write
Eq.~(\ref{enfunc}) as 
\begin{equation}
E[n] = T^{\prime}_{t_{\ast}}[n]+ \sum_{\bfrsub}\left(\frac{t_{\ast}}{T}\right)
         v_{\bfrsub}n_{\bfrsub} +
        \xi_{xc}[n] , 
\label{enfunc2}
\end{equation}
where 
\begin{displaymath}
\xi_{xc}[n]\equiv Q[n] - T^{\prime}_{t_{\ast}}[n] 
\end{displaymath}
is the exchange-correlation
energy functional. As noted in Section~\ref{sec1}, it is through the
exchange-correlation energy functional $\xi_{xc}[n]$ that DFT 
accounts for \textit{all many-body effects}. Since $n^{g}_{\bfrsub}$ 
minimizes both $\epsilon_{ks}[n]$ and $E[n]$, Eqs.~(\ref{ksenfunc}) and 
(\ref{enfunc2}) are stationary about $n=n^{g}$. Taking their functional 
derivatives with respect to $n$, evaluating the result at $n=n^{g}$, and 
eliminating $\delta T^{\prime}_{t_{\ast}}/\delta n|_{n=n^{g}}$ gives 
\begin{equation}
v^{ks}_{\bfrsub} = 
v_{\bfrsub} + \left(\frac{T}{t_{\ast}}\right)\, v_{xc}[n^{g}](\bfr ), 
\end{equation} 
for $t_{\ast}\neq 0$. Here $v_{xc}[n^{g}](\bfr )$ is the exchange-correlation 
potential which is the functional derivative of the exchange-correlation
energy functional $\xi_{xc}[n^{g}]$:
\begin{displaymath}
v_{xc}[n^{g}](\bfr )=\frac{\delta\xi_{xc}[n^{g}]}{\delta n^{g}_{\bfrsub}} .
\end{displaymath} 
This sets in 
place the formulas for a self-consistent calculation of the ground-state 
properties of $H_{\ast}=H(t_{\ast})$ using GS-DFT. Entanglement\cite{kap} and 
its links to quantum phase transitions\cite{lidr} have been studied using 
GS-DFT. 

\subsection{Time Dependent Density Functional Theory} 
\label{sec4.2}

Here we establish the Runge-Gross theorem \cite{r&g} for the instantaneous 
MAXCUT dynamics. Thus we focus on the instantaneous Hamiltonian $H_{\ast}=
H(t_{\ast})$ for a fixed $t_{\ast}$ ($0<t_{\ast} <T$). Now, however, we 
suppose that the external 
potential $v_{\bfrsub}$ in $H(t_{\ast})$ begins to vary at a moment we call 
$t=0$. For $t\leq 0$, $v_{\bfrsub}(t)= v_{\bfrsub}$, and the fermions are in 
the ground-state $|\psi_{0}\rangle$ of $H(t_{\ast})$. The Runge-Gross theorem 
states that the SOFs $n_{\bfrsub}(t)$ and $n^{\prime}_{\bfrsub}(t)$ evolving 
from a common initial state $|\psi (0) \rangle = |\psi_{0}\rangle$ under the 
influence of the respective potentials $V_{\bfrsub}(t)$ and 
$V^{\prime}_{\bfrsub}(t)$ (both Taylor-series expandable about $t=0$) will be 
different provided that $[ V_{\bfrsub}(t)- V^{\prime}_{\bfrsub}(t)]\neq C(t)$.
For us: 
\begin{eqnarray*}
V_{\bfrsub} (t)  & = &   \left(\frac{t_{\ast}}{T}\right)
         \left(1-\frac{t_{\ast}}{T}\right) v_{\bfrsub}(t) \\ 
V_{\bfrsub}^{\prime} (t) & = & \left(\frac{t_{\ast}}{T}\right)
       \left(1-\frac{t_{\ast}}{T}\right) v_{\bfrsub}^{\prime}(t)
\end{eqnarray*}
and
\begin{eqnarray*} 
V_{\bfrsub}(t) & = & \sum_{k=0}^{\infty} a_{k}(\bfr )t^{k}/k!;\\ 
V^{\prime}_{\bfrsub}(t) & = & \sum_{k=0}^{\infty} a^{\prime}_{k}(\bfr )
t^{k}/k!. 
\end{eqnarray*} 
Let $ C_{k}(\bfr ) \equiv a_{k}(\bfr ) - a^{\prime}_{k}(\bfr )$.
The condition that $[ V_{\bfrsub}(t)- V^{\prime}_{\bfrsub}(t)]\neq C(t)$ means 
a smallest integer $K$ exists such that $C_{k} (\bfr )$ is a non-trivial 
function of $\bfr$ for all $k\geq K$, while for $k<K$, it is a constant 
$C_{k}$ which can be set to zero without loss of generality. 

Recall (Eq.~(\ref{fermcurr})) that the conserved fermion current has components 
\begin{eqnarray*}
\hat{j}_{\bfrsub ,0} (t) & = & n_{\bfrsub}(t),\\ 
\hat{j}_{\bfrsub ,k}(t) & = & \left(\frac{1}{2\pi}\right)\sum_{\bfysub} 
  (\Delta_{k} G_{\bfr ,\bfy}) \partial_{t}n_{\bfysub}(t), 
\end{eqnarray*} 
with $k=1,2$. Defining $j_{\bfrsub ,k}(t) = \langle
\psi_{0}|\hat{j}_{\bfrsub ,k}(t)|\psi_{0}\rangle$, it follows that
\begin{equation}
{} \hspace{-0.05in}
\partial_{t}\{ j_{\bfrsub ,k}(t) - 
       j^{\prime}_{\bfrsub ,k}(t)\} = \langle\psi_{0}|[ 
      \hat{j\/}_{\bfrsub ,k}(t), H(t) - H^{\prime}(t)] |\psi_{0}\rangle .
\label{probeq}
\end{equation}
Here $j_{\bfrsub ,k}(t)$ ($j^{\prime}_{\bfrsub ,k}(t)$) and $H(t)$ 
($H^{\prime}(t)$) are the expected fermion current and the Hamiltonian,
respectively, when the external potential is $v_{\bfrsub}(t)$ 
($v^{\prime}_{\bfrsub}(t)$). The Hamiltonians $H(t)$ and $H^{\prime}(t)$
differ only in the external potential. Defining
\begin{displaymath}
\delta j_{\bfrsub ,k}(t) = 
j_{\bfrsub ,k}(t)-j^{\prime}_{\bfrsub ,k} (t),
\end{displaymath} 
and 
\begin{displaymath}
\delta V_{\bfysub}(t) = 
V_{\bfysub}(t)-V^{\prime}_{\bfysub}(t), 
\end{displaymath}
evaluation of the commutator in Eq.~(\ref{probeq}) eventually gives 
\begin{equation}
{} \hspace{-0.1in}
\partial_{t}\{ \delta j_{\bfrsub ,k}(t) \} = -\left(\frac{1}{2\pi}\right)
             \sum_{\bfysub}
         (\Delta_{k}G_{\bfr ,\bfy})\; \delta V_{\bfysub}(t) \; 
           \mathcal{M}_{\bfy}(t) ,
\label{currder}
\end{equation}
where 
\begin{displaymath}
\calM_{\bfysub}(t) = \langle\psi_{0}|(a^{\dagger}_{\bfysub}Q_{\bfysub}
+Q^{\dagger}_{\bfysub}a_{\bfysub})|\psi_{0}\rangle. 
\end{displaymath}
With $K$ defined as above,
taking $K$ time-derivatives of Eq.~(\ref{currder}) and evaluating the result 
at $t=0$ gives
\begin{equation}
{} \hspace{-0.15in}
\left. \frac{\partial^{K+1}}{\partial t^{K+1}}(\delta j_{\bfrsub ,k}(t))
   \right|_{0} =
  -\left(\frac{1}{2\pi}\right)\sum_{\bfysub} (\Delta_{k} G_{\bfrsub ,\bfysub})
    \calM_{\bfysub}(0) C_{K}(\bfr ) ,
\label{Kdervscurr}
\end{equation} 
where we have used that 
\begin{displaymath}
\frac{\partial^{k}}{\partial t^{k}}\left(\delta V_{\bfysub}(t)\right)
|_{t=0} = C_{k}(\bfy ) = 0 
\end{displaymath}
for $k<K$. It is important to note that 
$\calM_{\bfysub}(0)\neq 0$. This follows since 
\begin{displaymath}
[H(t_{\ast}), n_{\bfrsub}(
t_{\ast})] \neq 0 
\end{displaymath}
for $t_{\ast}\neq T$, 
and so the eigenstates of $H( t_{\ast})$ (specifically, its ground-state 
$|\psi_{0}\rangle$) cannot be fermion number eigenstates. This ensures that 
the ground-state expectation value 
\begin{displaymath}
\calM_{\bfysub}(0) =\langle\psi_{0}|(a^{\dagger}_{\bfysub}Q_{\bfysub}+
Q^{\dagger}_{\bfysub}a_{\bfysub})|\psi_{0}\rangle\neq 0 
\end{displaymath}
for $t_{\ast}\neq T$. It follows from the continuity
equation for the fermion current that 
\begin{displaymath}
\frac{\partial}{\partial t}\left(n_{\bfrsub}(t)-
n_{\bfrsub}^{\prime}(t)\right) = -\sum_{k=1,2}\Delta_{k}\:
        \left\{\delta j_{\bfrsub ,k}(t)\right\}.
\end{displaymath} 
Taking K time-derivatives of this equation, evaluating the result at $t=0$, 
and using Eq.~(\ref{Kdervscurr}) gives
\begin{equation}
\frac{\partial^{K+2}}{\partial t^{K+2}}\left( n_{\bfrsub}(t) - 
   n^{\prime}_{\bfrsub}(t)\right)|_{t=0} = - C_{K}(\bfr )\calM_{\bfrsub}(0) 
\neq 0 ,
\label{bigres}
\end{equation} 
where we have used the equation of motion for $G_{\bfrsub ,\bfysub}$.
Equation~(\ref{bigres}) indicates that $n_{\bfrsub}(t)$ cannot equal 
$n^{\prime}_{\bfrsub}(t)$ since it insures that they will be different at 
$t=0^{+}$, and so cannot be the same function. This proves the Runge-Gross 
theorem for the instantaneous MAXCUT dynamics. 

We have just seen that when potentials $V_{\bfrsub}(t)$ and 
$V^{\prime}_{\bfrsub}(t)$ differ by a time-dependent function $C(t)$, they  
give rise to the same SOF $n_{\bfrsub}(t)$. However, the wavefunctions
produced by these potentials from the same initial state will differ by a 
time-dependent phase factor. For our purposes, it is important to note that 
this extra phase factor cancels out when calculating the expectation value of 
an operator. In particular, it will cancel out when calculating the
instantaneous energy eigenvalues $E_{n}(t)=\langle E_{n}(t)|H(t)|E_{n}(t)
\rangle$. As a result, this phase factor will not affect our calculation 
of the minimum energy gap in Section~\ref{sec5}. Having said that, it is worth
noting  that this subtlety is not expected to cause difficulties in practice 
since the probe potential $V_{\bfrsub}(t)$ is assumed to be under the direct
control of the experimenter, and so the precise form of $V_{\bfrsub}(t)$ is 
\textit{known}. When an experimentalist says a sinusoidal probe potential has 
been applied, this means $V_{\bfrsub}(t)= V_{\bfrsub}\sin\omega t$; it
does \textit{not\/} mean $V_{\bfrsub}(t)=V_{\bfrsub}\sin\omega t + C(t)$. 
Thus in a well-designed experiment $C(t)=0$.

The Kohn-Sham (KS) system of 
non-interacting fermions can also be introduced in TD-DFT \cite{r&g}. We must 
still assume that the interacting SOF $n_{\bfrsub}(t)$ can be 
obtained from the SOF of the non-interacting KS fermions moving in the 
external potential $v^{ks}_{\bfrsub}(t)$. The potentials $v^{ks}_{\bfrsub}(t)$ 
and $v_{\bfrsub}(t)$ are related via ($t_{\ast}\neq 0$)
\begin{equation}
v^{ks}_{\bfrsub}(t) = v_{\bfrsub}(t) + \left(\frac{T}{t_{\ast}}\right)\, 
  v_{xc}[n(t)](\bfr ) 
\label{tdvvrel}
\end{equation} 
though Eq.~(\ref{tdvvrel}) is to be thought of as defining the time-dependent 
exchange-correlation potential $v_{xc}[n(t)](\bfr )$.

\section{Minimum Gap} 
\label{sec5}

A problem of longstanding treachery in GS-DFT is the calculation of the 
excitation energies of a fermion system. TD-DFT was able to find these 
energies by determining the system's frequency-dependent linear response, and 
relating the excitation energies to poles appearing in that response. The 
arguments used\cite{exen} are quite general, and can be easily adapted to 
determine the energy gap for the instantaneous MAXCUT dynamics. 

Previously, we considered an external potential that becomes time-varying for 
$t\geq 0$. Our interest is in the \textit{interacting\/} fermion linear 
response, and so we assume that the total potential has the form 
\begin{displaymath}
v^{\mathrm{tot}}_{\bfrsub}(t)= v_{\bfrsub} +v^{1}_{\bfrsub}(t), 
\end{displaymath}
with $v^{1}_{\bfrsub}(t)$ a suitably small time-varying perturbation. The probe 
potential $v^{1}_{\bfrsub}(t)$ generates a first-order response 
$n^{1}_{\bfrsub}(t)$ in the SOF: 
\begin{displaymath}
n^{\mathrm{tot}}_{\bfrsub}(t) = 
n^{g}_{\bfrsub}+n^{1}_{\bfrsub}(t). 
\end{displaymath}
The susceptibility $\chi_{\bfrsub , 
\bfrsub^{\prime}}(t-t^{\prime})$ connects the first-order probe potential to 
the  SOF response. The total potential $v_{\bfrsub}^{\mathrm{tot}}(t)$ is 
related to the KS potential $v^{ks}_{\bfrsub}(t)$ through Eq.~(\ref{tdvvrel}), 
and by assumption, the SOF for both the interacting and KS fermions is the 
same. This allows the time-Fourier transform of the SOF response 
$n^{1}_{\bfrsub}(\omega )$ to be determined from the time-Fourier transforms 
of the KS susceptibility $\chi^{ks}_{\bfrsub ,\bfrsub^{\prime}}(\omega )$, the 
exchange-correlation kernel $f_{xc}[n^{g}](\bfr ,\bfr^{\prime};\omega )$, and 
the probe potential $v^{1}_{\bfrsub}(\omega )$: 
\begin{eqnarray}
\lefteqn{{} \hspace{-0.3in} \sum_{\bfysub^{\prime}}\{
  \delta_{\bfrsub ,\bfysub^{\prime}} - 
 \sum_{\bfrsub^{\prime}}\chi^{ks}_{\bfrsub ,\bfrsub^{\prime}}(\omega )
  f_{xc}[n^{g}](\bfr^{\prime},\bfy^{\prime};\omega )\}
   n^{1}_{\bfysub^{\prime}}(\omega )   } \nonumber \\
 & & {} \hspace{1.3in}   = \sum_{\bfrsub^{\prime}}
           \chi^{ks}_{\bfrsub ,\bfrsub^{\prime}}(\omega )
     v^{1}_{\bfrsub^{\prime}}(\omega ) . 
\label{uglyeq}
\end{eqnarray}
The KS susceptibility \cite{m&g} depends on the KS static unperturbed orbitals 
$\phi^{j}_{\bfrsub}$; and the corresponding energy eigenvalues 
$\varepsilon_{j}$ and orbital occupation numbers $f_{j}$:
\begin{eqnarray}
{}\hspace{-0.25in}
\chi^{ks}_{\bfrsub ,\bfrsub^{\prime}}(\omega ) & = &
 \sum_{j,k}\left( f_{k}-f_{j}\right)
\frac{\phi_{j}(\bfr )\overline{\phi}_{k}(\bfr)\overline{\phi}_{j}
  (\bfr^{\prime})
          \phi_{k}(\bfr^{\prime})}{\omega -(\varepsilon_{j}-\varepsilon_{k})
            +i\eta}.
\end{eqnarray} 
The exchange-correlation kernel $f_{xc}[n^{g}]$ incorporates all many-body
effects into the linear response dynamics, and is related to the 
exchange-correlation potential $v_{xc}[n^{g}]$ through a functional derivative:
\begin{displaymath}
f_{xc}[n^{g}]=\frac{\delta v_{xc}[n^{g}]}{\delta n^{g}}.
\end{displaymath}
In general, the \textit{interacting\/} fermion excitation energies 
\begin{displaymath}
\Omega_{jk}=E_{j}- E_{k}
\end{displaymath} 
differ from the KS excitation energies 
\begin{displaymath}
\omega_{jk}=\varepsilon_{j}- \varepsilon_{k}.
\end{displaymath} 
The RHS of Eq.~(\ref{uglyeq}) 
remains finite as $\omega \rightarrow\Omega_{jk}$, while the first-order SOF 
response $n^{1}_{\bfysub^{\prime}}(\omega )$ has a pole at each $\Omega_{jk}$. 
Thus the operator on the LHS acting on $n^{1}_{\bfysub^{\prime}}(\omega )$ 
cannot be invertible. Otherwise, its inverse could be applied to both sides 
of Eq.~(\ref{uglyeq}) with the result that the RHS would remain finite as 
$\omega\rightarrow \Omega_{jk}$, while the LHS would diverge. To avoid this 
inconsistency, the operator must have a zero eigenvalue as $\omega\rightarrow
\Omega_{jk}$. Following Ref.~\onlinecite{exen}, one is led to the following 
eigenvalue problem:  
\begin{equation}
\sum_{k^{\prime},j^{\prime}}
  \frac{M_{kj;k^{\prime}j^{\prime}}(\omega )}{\omega - 
   \omega_{j^{\prime}k^{\prime}} + i\eta}\xi_{k^{\prime}j^{\prime}}(\omega )
  = \lambda (\omega )\xi_{kj}(\omega ),
\label{eigvalprob}
\end{equation}
where, writing 
\begin{displaymath}
\alpha_{k^{\prime}j^{\prime}}=f_{k^{\prime}}-f_{j^{\prime}}, 
\end{displaymath} 
and 
\begin{displaymath}
\Phi^{kj}_{\bfrsub} = \overline{\phi}_{k}(\bfr )\phi_{j}(\bfr ),
\end{displaymath}
we have
\begin{displaymath}
M_{k^{\prime}j^{\prime};kj}(\omega ) = \alpha_{k^{\prime}j^{\prime}}
 \sum_{\bfrsub^{\prime},\bfysub^{\prime}}\:
  \overline{\Phi}^{kj}_{\bfrsub^{\prime}}
  \:\left\{ f_{xc}[n^{g}](\bfr^{\prime},\bfy^{\prime};\omega )\right\}
 \:\Phi^{k^{\prime}j^{\prime}}_{\bfysub^{\prime}}(\omega ) .
\end{displaymath}
It can be shown that $\lambda (\Omega_{jk} ) = 1$. 

At this point in the argument, it proves necessary to introduce some form of 
approximation to proceed further. In the single-pole approximation\cite{exen} 
the KS poles are assumed to be well-separated so that we can focus on a 
particular KS excitation energy $\omega_{jk}=\omega_{\ast}$. The eigenvectors
$\xi_{k^{\prime}j^{\prime}}(\omega )$ and the matrix operator $M_{kj;k^{\prime}
j^{\prime}}(\omega )$ are finite at $\omega_{\ast}$, while the eigenvalue
$\lambda (\omega )$ must have a pole there to match the pole on
the LHS of Eq.~(\ref{eigvalprob}): 
\begin{displaymath}
\lambda (\omega ) = \frac{A(\omega_{\ast})}{(\omega -\omega_{\ast})} + 
  \mathcal{O}(1). 
\end{displaymath}
Let $\omega_{\ast}$ be $d$-fold 
degenerate: $\omega_{k_{1}j_{1}},\cdots ,\omega_{k_{d}j_{d}}= 
\omega_{\ast}$. Matching singularities in Eq.~(\ref{eigvalprob}) 
gives
\begin{equation}
\sum_{l=1}^{d} M_{k_{i}j_{i};k^{\prime}_{l}j^{\prime}_{l}}(\omega_{\ast})
 \;\xi^{n}_{k^{\prime}_{l}j^{\prime}_{l}} = A^{n}(\omega_{\ast})
  \;\xi^{n}_{k_{i}j_{i}}(\omega_{\ast}) ,
\label{secondeig}
\end{equation} 
where $i,n=1,\ldots ,d$. For our purposes, the eigenvalues $A^{n} 
(\omega_{\ast})$ are of primary interest and are found from 
Eq.~(\ref{secondeig}). From each $A^{n}(\omega_{\ast})$, we find
\begin{displaymath}
\lambda^{n} (\omega ) = \frac{A^{n}(\omega_{\ast})}{(\omega - \omega_{\ast})}. 
\end{displaymath}
Since 
$\lambda^{n}(\Omega_{jk}) = 1$, it follows that the sum of $\lambda^{n}
(\Omega_{jk})$ and its complex conjugate is $2$. Plugging into this sum the 
singular expressions for $\lambda^{n} (\Omega_{jk})$ and that of its complex 
conjugate, and solving for $\Omega_{jk}^{n}$ gives 
\begin{displaymath}
\Omega^{n}_{jk} = 
\omega_{\ast} + Re[A^{n}(\omega_{\ast})].
\end{displaymath}
 \textit{Interactions\/} will thus 
generally \textit{split\/} the $\omega_{\ast}$-degeneracy. Now let 
\begin{displaymath}
\delta E =
\min_{n} Re [A^{n}(\omega_{\ast})]
\end{displaymath} 
and 
\begin{displaymath}
\overline{\Omega}_{jk}=\min_{n} 
\Omega^{n}_{jk}.
\end{displaymath} 
Our expression for $\Omega^{n}_{jk}$ then gives
\begin{displaymath}
\overline{\Omega}_{jk} = \omega_{\ast} + \delta E. 
\end{displaymath}
In the context of the QAE 
algorithm, our interest is the energy gap 
\begin{displaymath}
\Delta (t_{\ast}) = 
E_{1}(t_{\ast})-E_{0}(t_{\ast})
\end{displaymath} 
separating the instantaneous 
ground and first-excited states. In this case, our expression for 
$\overline{\Omega}_{jk}$ gives
\begin{equation}
\Delta (t_{\ast}) = [\varepsilon_{1}(t_{\ast}) - \varepsilon_{0}(t_{\ast})] + 
\delta E(t_{\ast}).
\label{gapres}
\end{equation}
To obtain the minimum gap $\Delta$ for QAE numerically, one picks a 
sufficiently large number of $t_{\ast}\hspace{-0.25em}\in\hspace{-0.25em}
(0,T)$; solves for $\Delta (t_{\ast})$ using the KS system 
associated with $H(t_{\ast})$ to evaluate the RHS of Eq.~(\ref{gapres}); then 
uses the minimum of the resulting set of $\Delta (t_{\ast})$ to upper bound 
$\Delta$. Because the KS dynamics is non-interacting, it has been possible 
to treat KS systems with $N\sim 10^{3}$ KS fermions \cite{shimo,jiang,sanch}. 
This would allow the evaluation of the minimum gap $\Delta (N)$ for the QAE 
algorithm for $N\sim 10^{3}$.

\section{Discussion}
\label{sec6}

As with all KS calculations, the minimum gap calculation requires an
approximation for the exchange-correlation energy functional $\xi_{xc}[n]$.
Note that, because the qubits in a quantum register must be located at fixed 
positions for the register to function properly, the associated JW fermions 
are distinguishable since they are each pinned to a specific lattice site.
Consequently, anti-symmetrization of the fermion wavefunction is not required, 
with the result that the \textit{exchange energy vanishes\/} in the MAXCUT 
dynamics. The exchange-correlation energy functional $\xi_{xc}[n]$ is then 
determined solely by the correlation energy which can be calculated using the 
methods of Ref.~\onlinecite{CepAd}. Parametrization of 
these results yields analytical expressions for the correlation energy per 
particle which, upon differentiation, give $v_{xc}[n]$ 
and $f_{xc}[n]$. Replacing $n\rightarrow n_{\bfrsub}$ in $\xi_{xc}[n]$ gives 
the local density approximation (LDA) for GS-DFT; while $n\rightarrow
n_{\bfrsub}(t)$ gives the adiabatic local density approximation (ALDA) for 
TD-DFT. These simple approximations have proven to be remarkably successful,
and provide a good starting point for the minimum gap calculation. 
Self-interaction corrections to $\xi_{xc}[n]$ are not necessary since the 
two-fermion interaction [see Eq.~(\ref{fermHam})] has no self-interaction 
terms. Finally, because the fermions are pinned, it will be necessary to test 
the gap for sensitivity to derivative discontinuities\cite{bandG} in 
$\xi_{c}[n]$. 

\begin{acknowledgments}
We thank A. Satanin for stimulating conversations. F. Nori gratefully 
acknowledges partial support from the NSA, LPS, ARO, and NSF grant No.\ 
EIA-0130383; and F. Gaitan thanks T. Howell III for continued 
support.
\end{acknowledgments}

\end{document}